\documentclass[fleqn,usenatbib]{mnras}

\usepackage{newtxtext,newtxmath}
\usepackage[T1]{fontenc}
\DeclareRobustCommand{\VAN}[3]{#2}
\let\VANthebibliography\thebibliography
\def\thebibliography{\DeclareRobustCommand{\VAN}[3]{##3}\VANthebibliography}

\usepackage{graphicx}	
\usepackage{amsmath}	

\usepackage{xspace}
\usepackage[dvipsnames]{xcolor}
\usepackage{cancel}

\newcommand{\dd}{\mathrm{d}}
\newcommand{\Lya}{\,\ifmmode{{\mathrm{Ly}\alpha}}\else Ly$\alpha$\fi\xspace}
\newcommand{\HI}{{\text{H\MakeUppercase{\romannumeral 1}}}\xspace}
\newcommand{\cm}{\,\ifmmode{{\mathrm{cm}}}\else cm\fi}

\newcommand{\ergps}{\,{\rm erg}\,{\rm s}\ifmmode{}^{-1}\else${}^{-1}$\fi}
\newcommand{\Mpch}{\,{\rm Mpc}\,\ifmmode h^{-1}\else $h^{-1}$\fi}
\newcommand{\snru}{\,\ifmmode{\mathrm{Myr}^{-1}}\else Myr${}^{-1}$\fi}
\newcommand{\kms}{\,\ifmmode{\mathrm{km}\,\mathrm{s}^{-1}}\else km\,s${}^{-1}$\fi\xspace}

\title[Anisotropic Ly$\alpha$ escape]{Crossing walls and windows: the curious escape of Lyman-$\alpha$ photons through ionized channels
}

\author[Almada Monter \& Gronke]{
Silvia Almada Monter$^{1,2}$\thanks{E-mail: almada@mpa-garching.mpg.de}, Max Gronke$^{1}$
\\
$^{1}$Max Planck Institute for Astrophysics, Karl-Schwarzschild-Str. 1, D-85741 Garching, Germany\\
$^{2}$Ludwig-Maximilians-Universität München, Geschwister-Scholl-Platz 1, D-80539 München, Germany}

\date{Draft from \today}

\pubyear{2024}

\begin{document}
\label{firstpage}
\pagerange{\pageref{firstpage}--\pageref{lastpage}}
\maketitle

\begin{abstract}
The diverse Lyman-alpha (\Lya) line profiles are essential probes of gas in and around galaxies. While isotropic models can successfully reproduce a range of \Lya observables, the correspondence between the model and actual physical parameters remains uncertain. We investigate the effect of anisotropies on Ly$\alpha$ escape using a simplified setup: an empty hole (fractional size $\tilde s$) within a semi-infinite slab with constant column density. Due to the slab's high line-centre optical depth ($\tau_0\gtrsim 10^{5-6}$), most photons should escape through the empty channel.
However, our numerical findings indicate that only a fraction $\sim \tilde s$ of photons exit through this channel. To explain this puzzle, we developed an analytical model describing the scattering and transmission behaviour of \Lya photons in an externally illuminated slab. Our findings show that the number of scatterings per reflection follows a Lévy distribution ($\propto N^{-3/2}$). This means that the mean number of scatterings is orders of magnitude greater than expectations, facilitating a shift in frequency and the subsequent photon escape. Our results imply that \Lya photons are more prone to traverse high-density gas and are surprisingly less biased to the `path of least resistance'. Hence, \Lya can trace an average hydrogen distribution rather than only low-column density channels. 
\end{abstract}





\section{Introduction}

The Universe's brightest spectral line, Lyman-$\alpha$ (\Lya) plays a critical role in various fields of astronomy and astrophysics. Since the \Lya line profile is sensitive to the gas column density ~\citep{adams_escape_1972,neufeld_transfer_1990,dijkstra_ly_2006}, gas kinematics ~\citep{bonilha_monte_1979, ahn_ly_2002}, dust ~\citep{neufeld_transfer_1990, laursen_ly_2009}, and fragmentation~\citep{neufeld_escape_1991, hansen_lyman_2006,gronke_mirrors_2016}, it provides crucial information about the gas that photons traverse on their journey to the observer. Hence, \Lya has been an effective tool for studying the properties of the interstellar medium (ISM), the circumgalactic medium (CGM), and its spatial distribution ~\citep[e.g.,][]{steidel_structure_2010,wisotzki_extended_2016} as well as for exploring galaxies with the highest redshifts ~\citep{dijkstra_lyman_2014, barnes_ly-_2014,ouchi_observations_2020}. 

As \Lya is a resonant line undergoing complex radiative transfer (RT) effects, decoding the information imprinted in the line profile remains a complex and unsolved challenge. Most attempts so far have relied on isotropic models that succeed at reproducing the line profile using simple geometries ~\citep[e.g., ][]{verhamme_3d_2006, dijkstra_ly_2006}. However, these models still leave unclear how the fitted parameters relate to actual physical conditions (e.g., ~\citealt{orlitova_puzzling_2018,gronke_lyman-_2016,li_deciphering_2022}). Factually, astrophysical gas configurations are highly anisotropic. Such anisotropies play an influential role in shaping the \Lya profile. This is confirmed by studies of \Lya RT through varied geometries in simplified ~\citep[e.g.,][]{behrens_inclination_2014,zheng_anisotropic_2014,chang_first_2024} and more complex ~\citep[e.g.][]{smith_physics_2022, kim_small_2023} setups. A notable fitted parameter is the neutral gas column density ($N_{\rm HI}$). Given the multiphase and anisotropic nature of the gas, we anticipate realistic column densities to be distributed within a wide range of values. However, current models yield only one value of $N_{\rm HI}$, leaving unresolved whether this quantity represents the mean of the distribution or is skewed towards lower densities where photons are more easily transmitted.

Accounting for anisotropies is essential for understanding the ISM. Stellar feedback, for instance, significantly contributes to this anisotropy by generating channels filled with a hot gas phase ~\citep{cox_large-scale_1974, mckee_theory_1977}. These low-density channels should also be imprinted in the \Lya line and offer a prime opportunity for radiation to escape and reach the observer ~\citep{zackrisson_spectral_2013}. It is commonly asserted that within a porous ISM \Lya primarily probes the gas phases with the lowest column density, essentially tracing the `path of the least resistance' ~\citep{jaskot_new_2019,kakiichi_lyman_2021}. However, the extent to which other phases in the distribution contribute remains uncertain. 

These low $N_{\rm HI}$ channels are of particular importance as their presence might 
 be crucial to allow ionizing photons to escape their host galaxy and to re-ionize the Universe \citep{zackrisson_spectral_2013, rivera-thorsen_sunburst_2017,herenz_vltmuse_2017, bik_super_2018}. Since \Lya and Lyman Continuum (LyC) probe the same neutral hydrogen, \Lya is a valuable tool for inferring ionizing escape fractions and understanding LyC escape \citep{verhamme_using_2015,dijkstra_ly-lyc_2016, izotov_low-redshift_2018}. However, thus far it is unclear how this proxy for LyC escape is influenced by an anisotropic gas distribution, that is, whether \Lya correlates with a `line-of-sight' LyC escape fraction or a `global one' - important for models of the epoch of reionization.

Given this inherently anisotropic nature of realistic neutral hydrogen distributions in and around galaxies, there is a growing need for a model that accounts for diverse gas distributions. Furthermore, no firm theoretical understanding of anisotropic \Lya escape has been established. In this letter, we first introduce the problem of anisotropic \Lya escape on a simplified, theoretical basis (\S~\ref{sec:analitycal_considerations}).
Then, we test our theoretical model in \S~\ref{sec:numerical} using results of Monte Carlo RT simulations before we discuss these results within a broader framework in \S~\ref{sec:Results}. We discuss and conclude in \S~\ref{sec:Discussion}.
\vspace{-0.5cm}
\section{Analytical considerations}
\label{sec:analitycal_considerations}
\subsection{Setup and first estimate}
\label{subsec:first_estimate}
\begin{figure}
    \centering   \includegraphics[width=0.99\columnwidth,keepaspectratio]{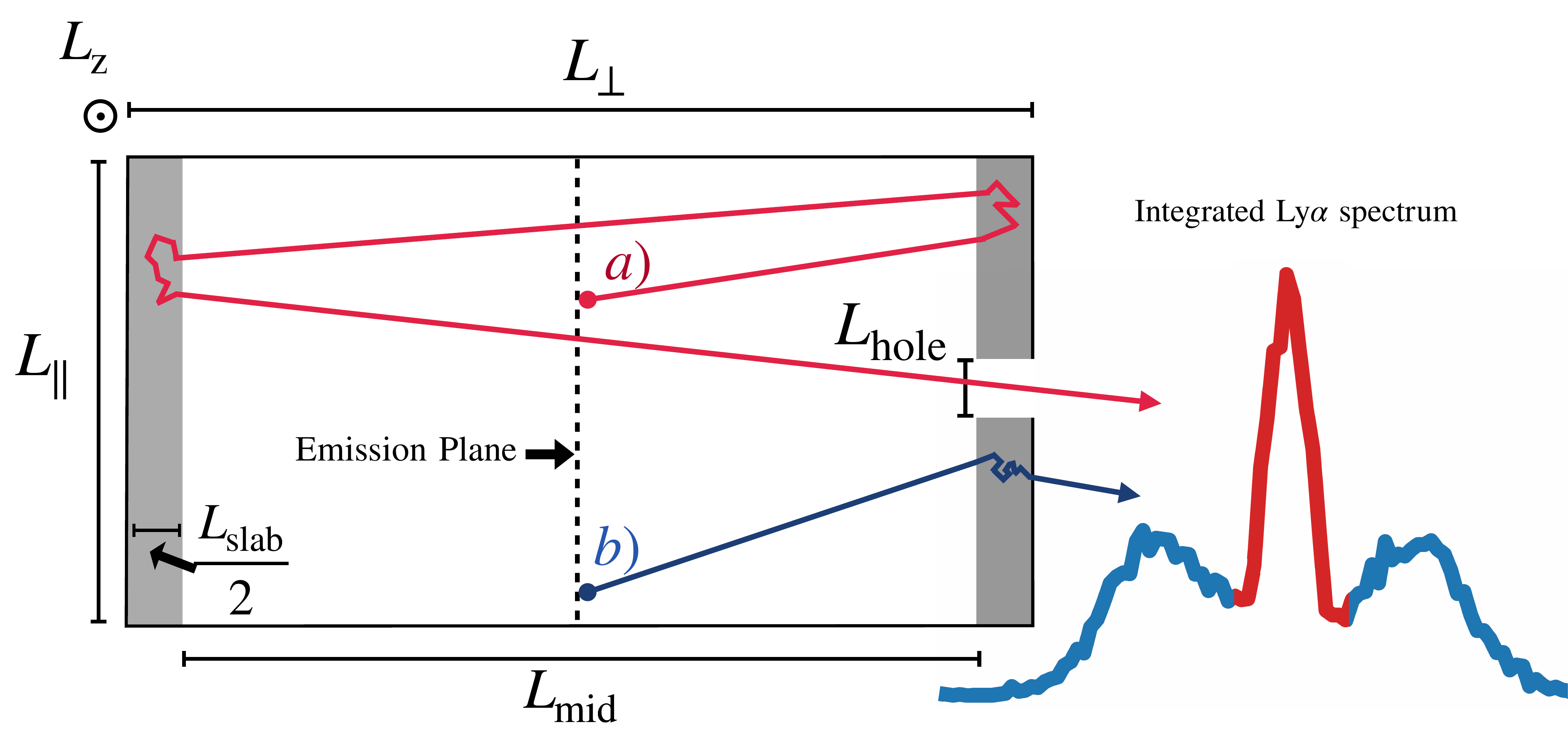}
    \caption{Simulation setup to investigate the impact of non-homogeneous gas distributions on the \Lya profile. Anisotropy is introduced by incorporating a square hole with a side length of $L_{\rm hole}$ along the non-periodic direction (see \S~\ref{sec:numerical} for more information). The arrow marked as (a) shows the path followed by an example photon that was reflected and bounced inside the slab before encountering the hole. Arrow (b) illustrates the path of a photon directly transmitted through the slab. The colour coding of the \Lya line shown on the right indicates the origin of the spectrum components}
    \label{fig:diagram}   
\vspace{-0.5 cm}
\end{figure}

In order to understand 
the escape of \Lya through an anisotropic medium let us consider the simple case of a static, semi-infinite slab with column density $N_{\HI}$ and temperature $T$ in which an empty hole is carved on one side with a surface area fraction $\tilde s$. This setup can be seen as a simplification of the complex distribution of hydrogen column densities of a porous ISM, as described above.
All the \Lya photons are emitted near the line centre in the mid-plane, where a cavity allows the initial diffusion to escape from the hole. Fig.~\ref{fig:diagram} shows a sketch of the considered setup.

The question we want to answer is which fraction of the flux will escape through the hole and which fraction will escape through the slab, to do so, we define the ratio $\tilde f = F_{\rm hole}/F_{\rm slab}$.
To first order, we can infer that the photons escaping through the slab form a broad double-peaked spectrum \citep{adams_escape_1972, neufeld_transfer_1990}. This characteristic double-peaked profile arises from Doppler shifts caused by the scattering events involving photons and atoms in thermal motion within the optically thick medium. In contrast, due to the lack of scattering material inside the hole, we infer that the photons escaping through it have a frequency close to the line centre, forming a central peak. Thus, $\tilde f$ will set the emergent \Lya spectrum (cf. illustration in Fig.~\ref{fig:diagram}). Using RT simulations, we confirm in \S~\ref{sec:Results} that this picture is not far from the truth. 

A very naive first estimate would be that $F_{\rm hole}\propto \tilde s$ and $F_{\rm slab}\propto \exp(-\tau_{\rm i})$ where $\tau_{\rm i}=N_\HI \sigma(x=x_i)$ is the optical depth at the emitted frequency $x_{\rm i}= (\nu-\nu_0)/\Delta\nu$, expressed here in Doppler shifts from line centre frequency $\nu_0$ (with $\Delta \nu = c (\nu / \nu_0-1)/v_{\rm th}$ and $v_{\rm th}$ is the thermal velocity). This estimate would yield, with $\tilde s\sim 0.1$ and $N_\HI=10^{19}\cm^{-2}$, that essentially all the photons would escape through the hole, i.e., $\tilde f\rightarrow\infty$, since $\tau(x=x_{\rm i}=0)\approx 6\times 10^5$ at $T=10^4\ $K. This approximation is clearly an oversimplification since the $\exp(-\tau_{\rm i})$ term does not include the resonant nature of \Lya. However, it illustrates the basic property of this setup, namely that for \Lya photons, it should be \textit{significantly} easier to escape through the hole than through the slab.

\citet{neufeld_transfer_1990} analytically solved the \Lya RT equation for a static, homogeneous slab, yielding a functional expression for the emergent spectrum alongside several additional properties. He found, for instance, that the transmission probability of incoming radiation is $T_{\rm slab}=4/(3 \tau_{\rm i})$. Thus, it would take $N_{\rm sc}^{\rm slab}= 1/T_{\rm slab}\sim \tau_{\rm i}$ scatterings to escape through the slab. During these scatterings, the \Lya photons can bounce off the walls of the cavity, and for each reflection, they have a probability of $\tilde s$ to escape through the hole. Hence, this improved estimate would yield $\tilde f\sim \tilde s/T_{\rm slab}\sim \tilde s \tau_{\rm i}$ which for the above-used values would equal $\sim 10^4$. Still,  nearly all the flux should pass through the hole, forming a centrally peaked spectrum. 

This picture leaves several problems. Firstly, this seems to be contradicting observations. Although the ISM is porous \citep{mckee_theory_1977}, we do not observe many centrally peaked \Lya spectra as most are double-peaked or have a redshifted single peak \citep[e.g.,][]{erb_ly_2014, yang_green_2016, hayes_spectral_2021}. In addition, for the few objects in which a triple peak is present \citep{dahle_discovery_2016,rivera-thorsen_sunburst_2017,rivera-thorsen_gravitational_2019,vanzella_ionising_2020}, the flux ratio of the outer to the inner peaks is $\mathcal{O}(1)$ but the peak separation of the outer peaks indicates column densities of $\sim 10^{19}\cm^{-2}$ \citep[cf.][]{rivera-thorsen_sunburst_2017}, implying a very small, highly fine-tuned $\tilde s$. The major problem with the above picture, however, is also that RT simulations of anisotropic \Lya escape as the ones we will present in \S~\ref{sec:Results} indicate much smaller central peak fluxes around $\tilde f\sim \mathcal{O}(1)$ \cite[see also numerical results by][who performed a range of \Lya radiative transfer simulations through anisotropic, outflowing media]{behrens_inclination_2014}.

\begin{figure*}
    \centering
    \includegraphics[width=2.2\columnwidth, keepaspectratio]{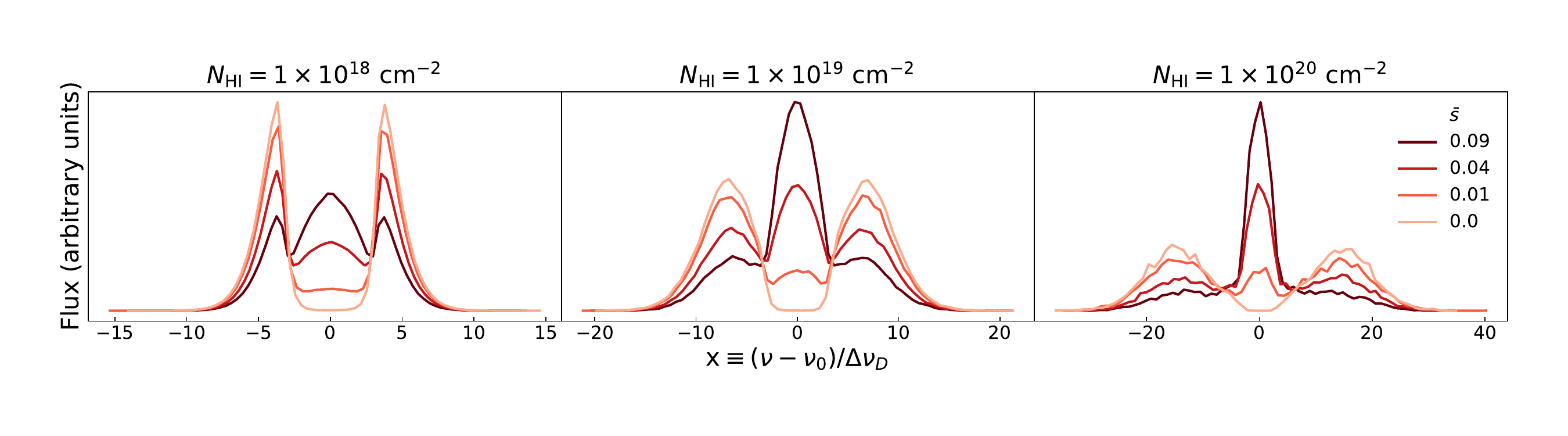}
    \vspace{-4em}
    \caption{\Lya spectra obtained from gas-filled slabs with an empty hole at varying $N_{\rm HI}$. The $\tilde s = 0$ case corresponds to a homogeneous slab without a hole (showing no central flux). The rest of the curves represent spectra with empty channels of different sizes. The dimensionless photon frequency is displayed on the x-axis. }
    \label{fig:example_spec}
\end{figure*}
\vspace{-0.5 cm}
\subsection{A novel picture of \Lya photon escape}
\label{sec:new_approach}
The problem with the above is that we assumed that while only a few of the photons are directly transmitted, the majority are almost immediately reflected by ‘bouncing' off the optically thick wall of the slab as if they were ping-pong balls. This simple transmission/reflection scenario suggests that the reflected photon suffers only a few scattering events per reflection; in other words, the number of scatters and reflections is roughly equal. However, a more realistic scenario should consider that the photon penetrates the slab and scatters several times before being reflected (i.e., the number of scatterings could be much greater than the reflections). Since every scattering offers the photon a chance to shift in frequency and physical space enough to escape through the slab, this larger number of scatterings significantly increases the photon's chances of being transmitted. While it is tempting to assume that due to the photons' difficulty penetrating the slab, they scatter only a few times per reflection, let us first compute this number analytically. To do so, consider a simple slab filled with neutral hydrogen with optical depth $\tau_0$, placed parallel to the xz plane, which is externally illuminated from one side (see \S~\ref{sec:numerical} for further description). 

This problem is analogous to the `Gambler's ruin' problem: a gambler starts in the casino with an initial sum of $w_0=\$1$ and has a $p=1/2$ probability of winning/losing $\$1$. Similarly,the \Lya photon enters the slab with some direction component parallel to the y-axis, a depth $\sim \lambda_{\rm mfp}=\left[\sigma_{\alpha} n_\HI\right]^{-1}$ (where $\sigma_{\alpha}$ is \Lya  absorption cross-section) and will move forward / backwards $\sim \lambda_{\rm mfp}$ (ignoring frequency redistribution). 

As the game progresses with time, the gambler's probability of success evolves. Analogously, the \Lya photons diffuse into the gas with a time-dependent probability $p(t,y)$ described by the zero-drift Fokker-Planck equation,
\begin{equation}\label{eq:difussion}
\frac{\partial p}{\partial t}=\frac{\sigma^2}{2}\frac{\partial^2 p}{\partial y^2},
\end{equation}
with $p(0,y) = \delta({y-y_0})$ as initial condition, diffusion coefficient $\sigma^2=\lambda_{\rm mfp}c$, and time $t$ or number of scatterings $N_{\rm scatter}=t c / \lambda_{\rm mfp}$ used interchangeably.

If we consider the edge of the slab as an absorbing barrier located at $y_{\rm b}=0$, we have the additional constraint $p(t,y=0) = 0$. This boundary value problem can be solved using
the initial conditions $p(0,y) = \delta(y-y_0) - A \delta(y-B)$ for which $A$ and $B$ are chosen so that the two evolving peaks cancel each other exactly at $y=y_{\rm b}$, i.e., $p(t,y_{\rm b})=0$ is fulfilled.
The solution of the diffusion equation with these initial conditions is
due to the linearity of the problem, simply two widening Gaussians with variance $\sigma^2 t$ and means $\pm y_0$. 
 To obtain the mean number of scatterings per reflection, let's consider the cumulative fraction of photons that have been reflected at time $t$:
\begin{equation}
    F_{\rm reflected}(t) = 1 - \int\limits_{0}^{\infty} p(t,y)\mathrm{d}y  = 1 - \mathrm{erf}\left(\frac{y_0}{\sqrt{2t}\sigma}\right)
     \label{eq:Freflected}
\end{equation}
where $\mathrm{erf}$ is the error function. Thus, the ones that are reflected within $[t\pm {\rm d}t/2]$ are given by
\begin{align}
    f_{\rm reflected}(t) = \frac{\mathrm{d}F_{\rm reflected}}{\mathrm{d}t} =& \frac{y_0}{\sqrt{2 \pi t^3 \sigma^2}}\exp\left(-\frac{y_0^2}{2 t \sigma^2}\right)\\
    \text{or  } f_{\rm reflected}(N_{\rm scatter})=&\frac{\mathrm{d}F_{\rm reflected}}{\mathrm{d}t}\frac{\mathrm{d}t}{\mathrm{d}N_{\rm scatter}}\nonumber\\=&\frac{1}{\sqrt{2 \pi N_{\rm scatter}^3}}\exp\left(-\frac{1}
    {2 N_{\rm scatter}}\right)
    \label{eq:f_reflected}
\end{align}
which is a Lévy distribution with scale $c_{\rm L}=1$ and location $\mu_{\rm L} = 0$. Note that for large number of scatterings, we obtain $f_{\rm reflected}\propto N_{\rm scatter}^{-3/2}$. 

Thus, the mean number of scatterings per reflection is simply
\begin{equation}
  \bar N_{\rm scatter} = \int\limits_0^{\mathrm{C}\tau_0} n f_{\rm reflected}(n) \dd n = \sqrt{\frac{2\mathrm{C}\tau_0}{\pi}}.
  \label{eq:nsc_mean}
\end{equation}
We have used that on average $\mathrm{C}\tau_0$  scattering events (with $\mathrm{C}\simeq2.66$) are required for the photon to shift to an escape frequency \citep{adams_escape_1972}. The above equations are the more general case of the `Gambler's ruin' problem described above. Eq.~\eqref{eq:Freflected} shows that eventually all gamblers are ruined ($F_{\rm reflected}\rightarrow 1$ as $t\rightarrow \infty$) but on average they stay infinitely long in the casino ($\bar N_{\rm scatter}\rightarrow\infty$ for $\tau_0\rightarrow\infty$; cf. Eq.~\ref{eq:nsc_mean}). However, \Lya photons are saved from this destiny due to their eventual shift in frequency to $x\gtrsim x_{\rm esc}$, where $x_{\rm esc}$ is a frequency that allows for escape \citep{adams_escape_1972,neufeld_transfer_1990}.

An additional result one can obtain from the above considerations is that the transmission probability through a slab is given by
\begin{align}
    T_{\rm slab}=&\alpha\int_{C\tau_0}^{\infty}f_{\rm reflected}(n) \dd n 
    \approx (2 \pi\mathrm{C}\tau_0)^{-1/2}.
    \label{eq:Tslab}
\end{align}
The additional factor of  $\alpha$ in Eq. \ref{eq:Tslab} arises because even once $x\gtrsim x_{\rm esc}$, the photon can escape forward through the slab or return backwards. Thus, if $\tau(x_{\rm esc})\lesssim 1$ (escape via single flight; cf. e.g. § 2.2 and § 2.3.3 in \citealp{gronke_resonant_2017} for a discussion on the different escape mechanisms through a homogeneous slab), these final escape paths are equally likely and $\alpha=1/2$. However, for very large optical depths, $\tau(x_{\rm esc})\gg1$, the photon needs to scatter a few more times to escape (via excursion) and will, in the process, undergo a second `Gamblers ruin' problem with $x\approx x_{\rm esc}$. The photon will then have a second chance to `win sufficient money to leave', i.e., reach the far end. In that case, $\alpha\approx 4/(3 \tau(x_{\rm esc}))$ following the original \citet{neufeld_transfer_1990} argument. Since this regime only applies for $a_{\rm v}\tau_0 > 1000$, i.e., very low temperatures or high optical depths, we will explore this regime in future work and here simply use $\alpha = 1/2$ (as in the last step in Eq.~\ref{eq:Tslab}).

With these results at hand, we can now estimate the ratio of photons escaping through the hole to those escaping through the slab as
\begin{equation}
  \label{eq:ftilde2}
  \tilde f\approx \frac{N_{\rm scatter}^{\rm slab} / \bar N_{\rm scatter}}{N_{\rm refl}^{\rm hole}}\approx \frac{\tilde s / 2}{T_{\rm slab}\bar N_{\rm scatter}}= \frac{\pi}{2}\tilde s
\end{equation}
where $N_{\rm refl}^{\rm hole}\simeq 2/\tilde s$ is the number of \textit{reflections} needed to escape through the hole.
Equation~\eqref{eq:ftilde2} shows two interesting things: \textit{(i)} a simple $\tilde f\sim \tilde s$ behaviour leading to a relatively low escape fraction through the hole (more in line with the observations discussed above), and \textit{(ii)} no dependence on $\tau_0$ which seems counter-intuitive as one might think of \Lya escaping preferentially through low column density channels. Note that although these results were derived using a slab as the main geometry choice, we expect these to hold for spherical geometries, since it has been shown that a slab captures the main RT effects, which would also emerge in a spherical setup~\citep[e.g.,][]{neufeld_transfer_1990, dijkstra_ly_2006}. Additionally, in our setup, the empty space around the emission plane is large enough to ensure a uniform probability of hitting the hole, simplifying the problem. This behaviour is expected to naturally occur in a spherical setup, where photons can hit the hole, even without extreme reflection angles. We will show this numerically in future work. In \S~\ref{sec:Results}, we compare these analytical estimates to our numerical findings.
\vspace{-0.5 cm}
\section{Numerical Method} 
\label{sec:numerical}
To test the theory presented in the previous section, we conducted \Lya RT simulations using the Monte Carlo RT code \texttt{tlac} \citep{gronke_directional_2014} which follows \Lya photon packages in real and frequency space taking into account scattering with \HI \citep[see, e.g.,][for an explanation of the technique]{dijkstra_lyman_2014}. We generated two different groups of numerical simulations: 
\begin{enumerate}
\item \textit{Semi-infinite slab with an empty hole.} The first group is to investigate the impact of non-homogeneous gas distributions on the \Lya profile by introducing anisotropies into the setup. Specifically, we aim to compare the theory developed in \S~\ref{sec:analitycal_considerations} to RT simulations. These simulations were conducted on a semi-infinite slab of size $L_{\parallel}\times L_{\perp}\times L_{\rm z}$ where the sides $L_{\parallel}$ and $L_z$ are periodic and $L_{\parallel}=L_{\rm  z}$. We placed the emission plane at the mid-point relative to the finite direction (i.e. at $L_{\perp}/2$) and set up the grid such that $L_{\perp} = L_{\rm mid} + L_{\rm slab}$ with an empty region of width $L_{\rm mid}$ and two thin HI layers of total width $L_{\rm slab}$ located at the extremes. To introduce anisotropy, we carve a square hole orthogonal to the emission plane with a side length $L_{\rm hole}$, as illustrated by Fig.~\ref{fig:diagram}. Note that the hole is consistent with the slab's periodicity and repeats in both the $z$ and $\parallel$ directions. This results in an opening area of $\tilde s \equiv L_{\rm hole}^2 / (L_{\parallel}L{\rm z})$. In this set of simulations, we maintained the gas temperature at $T=10^4$ K. In the derivation of \S~\ref{sec:analitycal_considerations}, the slab is assumed to be infinitely thin, and the probability of hitting the hole is independent of the reflecting position on the far side of the setup. These conditions are fulfilled if $L_{\rm slab} \ll L_{\rm hole}$ and $L_{\rm mid} \gg L_{\parallel}$, respectively. We thus chose $L_{\perp}=50 \ L_{\parallel}$ and $L_{\rm slab}=4\times 10^{-5}\ L_{\perp}$. For this study, we validated that changes in these parameters, within a factor of a few, do not affect our results. We will expand to a spherical setup and explore other effects of geometry in a follow-up paper.
Note also that since we are mainly interested in the interaction of photons with the gas inside the slab system until they find their way out either through the hole or through the slabs, our model omits the photons that initially escaped without interaction (e.g., no scattering events). As we show and discuss in Appendix~\ref{app:nointeraction} this contribution is $\frac{1}{2}\tilde s$.
\label{item:hole-setup}

\item \textit{Externally illuminated slab.} The second group of simulations is to test how many scatterings occur per reflection.
We created a slab externally illuminated from one side by positioning the photon emission plane at the bottom  of the slab. This setup does not include a hole. Here, photons either traverse the slab entirely to the opposite side and are \textit{transmitted} or penetrate the slab and scatter freely before being \textit{reflected}. We conducted these at three different temperatures $T=10^2$, $10^4$ and $10^6$ K.\label{item:one-side-setup}
\end{enumerate}
The slabs in both groups of simulations contain static neutral hydrogen with a fixed column density (from the emission plane to the boundary) $N_{\rm HI} \in [10^{17}, 10^{20}]\ \cm^{-2}$. We generally used $10^4$- $10^6$ photon packages to obtain a converged result and used $100\times 50000 \times 100$ cells in order to resolve the hole as well as the \HI layer.

\begin{figure}
    \centering    \includegraphics[width=0.992\columnwidth, keepaspectratio]{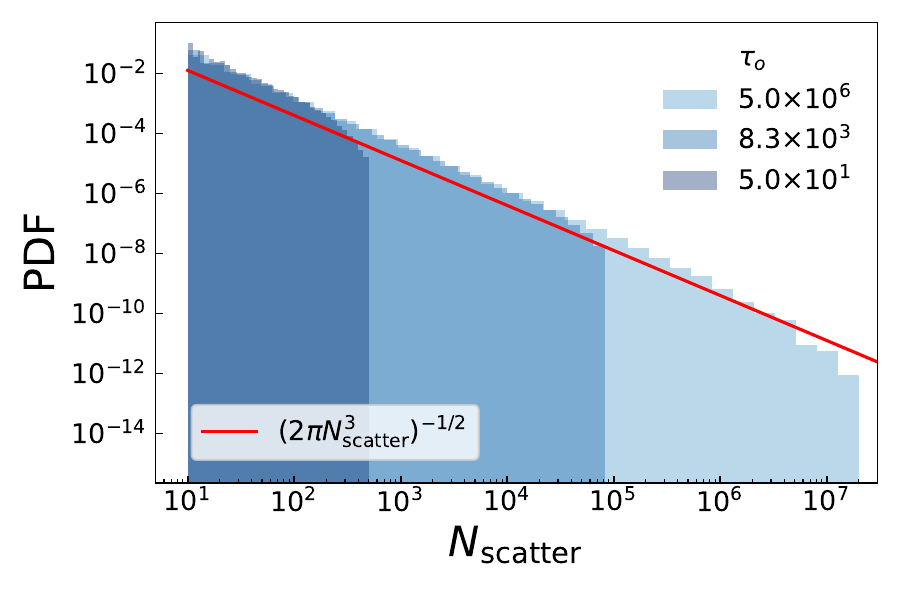}
    \vspace{-1.5em}
    \caption{Distribution of the scattering number for an externally illuminated slab for three selected values of optical depth. The solid line shows the analytical form derived from Eq.~\ref{eq:f_reflected}.}
    \label{fig:PDF}
    \vspace{-0.45cm}
\end{figure}
\vspace{-0.5cm}
\section{Numerical Results}
\label{sec:Results}

\subsection{\Lya spectra of a slab with a hole}
\label{sec:lyahole}

Fig.~\ref{fig:example_spec} shows the emergent \Lya spectra obtained from RT simulations as described in \S~\ref{sec:numerical}\ref{item:hole-setup}, at three different column density values. Each panel presents spectra from slabs with carved holes of various sizes $\tilde s$, including a homogeneous slab ($\tilde s=0$). At first glance, spectra from the homogeneous slab show the expected blue- and red-shifted peaks resulting from frequency diffusion. In contrast, spectra from a slab with a carved hole show flux near the line centre on top of the expected blue and red peaks.

Take, for instance, the last panel in Fig.~\ref{fig:example_spec} with $N_{\rm HI}= 10^{20}\ \rm cm^{-2}$ and $\tau_0\sim 6\times 10^6$. According to the discussion in \S~\ref{subsec:first_estimate}, for all values of $\tilde s$ in this figure, the ratio $\tilde f \simeq \tilde s \tau_0$ should indicate a central flux at least four orders of magnitude greater than the one escaping from the slab. In other words, most photons should exit through the hole, resulting in a single-peaked $\Lya$ line profile near the line centre. Evidently, the spectra in Fig.~\ref{fig:example_spec} contradict this prediction, as they show central fluxes with roughly the same order of magnitude as the red and blue peaks. We will further explore this central peak inconsistency  in \S~\ref{subsec:central_inconsistency}.
\vspace{-0.3 cm}
\subsection{Transmission and scattering statistics of an externally illuminated slab}
\label{subsec:testing}

The alternative theory outlined in \S~\ref{sec:new_approach} proposes that \Lya photons undergo a large number of scatterings before being reflected by the slab. To test this hypothesis, we examined the scattering behaviour of photons in a slab with an externally illuminated slab as described in \S~\ref{sec:numerical}\ref{item:one-side-setup}. In such a setting, photons can be either transmitted to the other side or reflected by the slab. Due to the medium's high optical depth, the majority of the photons get reflected upon encountering the slab. These reflected photons do not simply bounce/reflect as in a mirror but penetrate a small portion of the slab instead and `random walk' their way out. Consequently, the number of scatterings per reflection ($N_{\rm scatter}$) is significantly larger than the `simple reflection' scenario (where $N_{\rm scatter}=1$).

\begin{figure}
    \centering   \includegraphics[width=0.992\columnwidth, keepaspectratio]{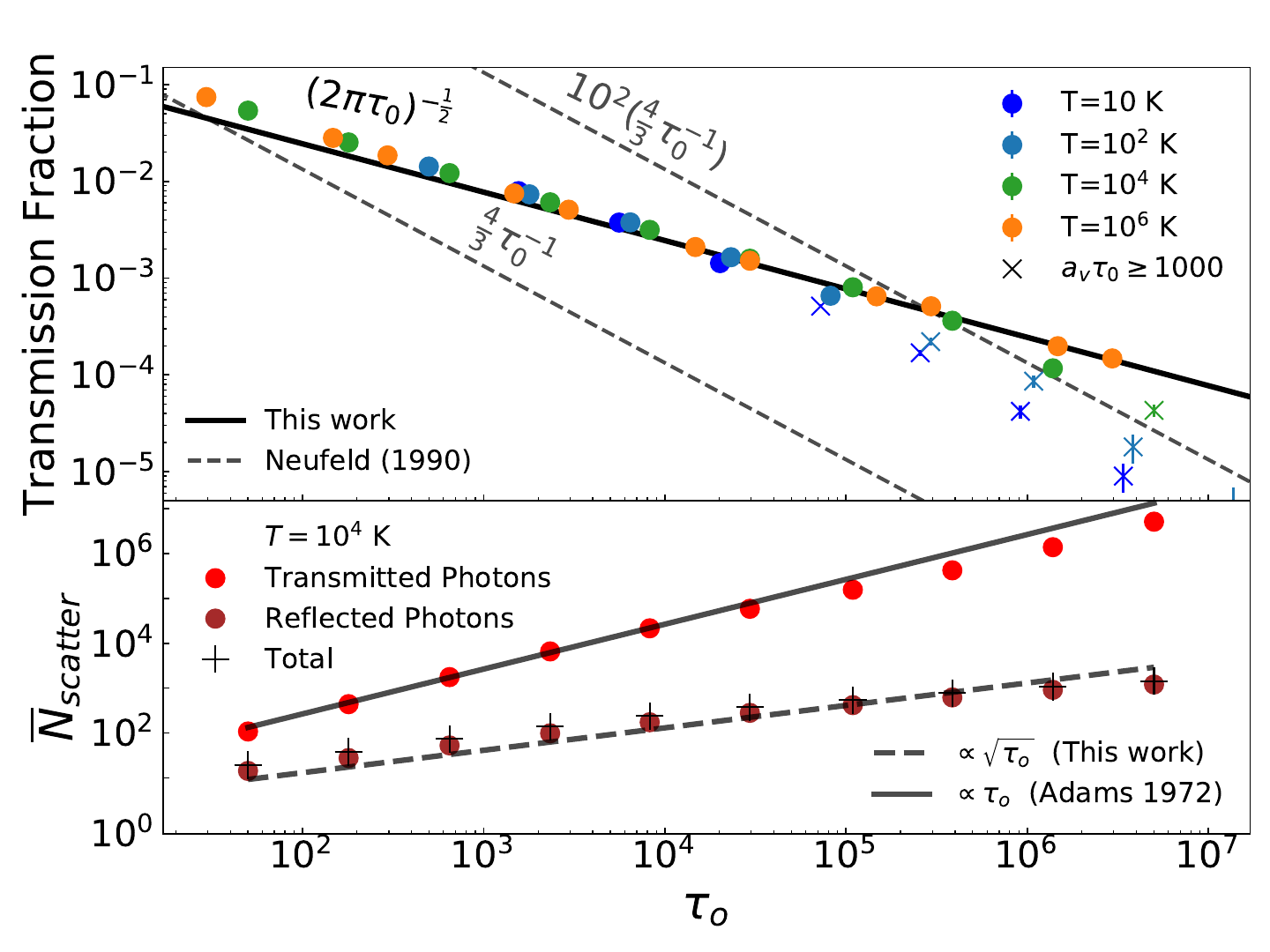}
    \caption{Transmission fraction as a function of $\tau_0$ through an externally illuminated homogeneous slab. Simulations are shown for different temperatures and compared with the analytical results. Bottom: mean number of scatterings of transmitted and reflected photons from an externally illuminated slab as a function of optical depth at $T=10^4$ K. Solid and dashed lines display the analytical solutions. The behaviour of the entire set of photons is represented with x-markers.}
    \label{fig:tau}
\end{figure}

Fig.~\ref{fig:PDF} displays the distribution of the number of scatterings per reflection $N_{\rm scatter}$ for photons reflected by slabs with three selected line centre optical depths at $T=10^4$ K. As anticipated by Eq.~\ref{eq:f_reflected}, this approximates a Lévy distribution, which for large values of $N_{\rm scatter}$, simplifies to a power law with an exponent of $-3/2$ (marked with a solid line in Fig.~\ref{fig:PDF}). Accordingly, photons are expected to undergo an average $\bar N_{\rm scatter}\propto \sqrt{\tau_0} \ $ scatterings (as given by Eq.~\ref{eq:nsc_mean}) before being \textit{reflected}. The \textit{transmitted} photons, on the other hand, require a mean number of scatterings $\propto \tau_0$ (as proposed by \citealt{adams_escape_1972}) to shift in frequency and increase the mean free path sufficiently and escape in a single excursion. These expectations are confirmed in the bottom panel of Fig.~\ref{fig:tau}, where we show the agreement between the simulated photons and their respective theoretical predictions across a wide range of optical depths. Only a minor fraction of the total photons can traverse the entire slab and exit on the opposite side (i.e., are transmitted). Eq.~\ref{eq:Tslab} suggests a new transmission probability approach where $T \sim \tau_0^{-1/2}$ instead of $T \sim \tau_0^{-1}$ \cite{neufeld_transfer_1990}. We performed simulations at different gas temperatures to assess the validity of both the new and previous approaches across a range of optical depth regimes. These results are displayed in the top panel of Fig.~\ref{fig:tau}. In general, the simulated transmission fraction aligns more closely with the new approach regardless of temperature except at the extremely large optical depth regime $a_v \tau_0\gtrsim 1000$ (with the Voigt parameter $a_v=4.7\times10^{-4}$ [$T/10^{4}\ \rm{K}$]$^{-1/2}$). At $T=10^4$ K, for instance,  the transmitted fraction is up to 2.5 orders of magnitude larger than the expected by \cite{neufeld_transfer_1990}. The temperature independence is consistent with \cite{adams_escape_1972}, who concluded that the probability of escape is independent of $a_v$ and temperature.

We have identified the extremely large optical depth regime in Fig.~\ref{fig:tau}. At this regime, simulations show a deviation from the $\propto \tau_0^{-1/2}$ trend. Interestingly,  the transmission agrees with the considerations presented in \cite{neufeld_transfer_1990}, i.e., $\propto \tau_0^{-1}$. However, such extreme optical depths or low temperatures are uncommon in astrophysical HI environments. Furthermore, for $T\lesssim 100\,$K, the frequency change due to recoil (which we ignore here) is non-negligible. Nevertheless, we will explore these deviations in future work.
\vspace{-0.3 cm}
\subsection{Photon escape through an ionized hole}
\label{subsec:central_inconsistency}

Given that the number of scatterings per reflection is much larger than initially anticipated, it becomes a relevant factor to consider when modelling the emergent \Lya spectra (such as the ones presented in Fig.~\ref{fig:example_spec}) from anisotropic settings. In particular, we aim to understand the effects of low-density channels and the central peak inconsistency described at the beginning of this section.

\begin{figure}
    \centering
    \includegraphics[width=0.992\columnwidth, keepaspectratio]{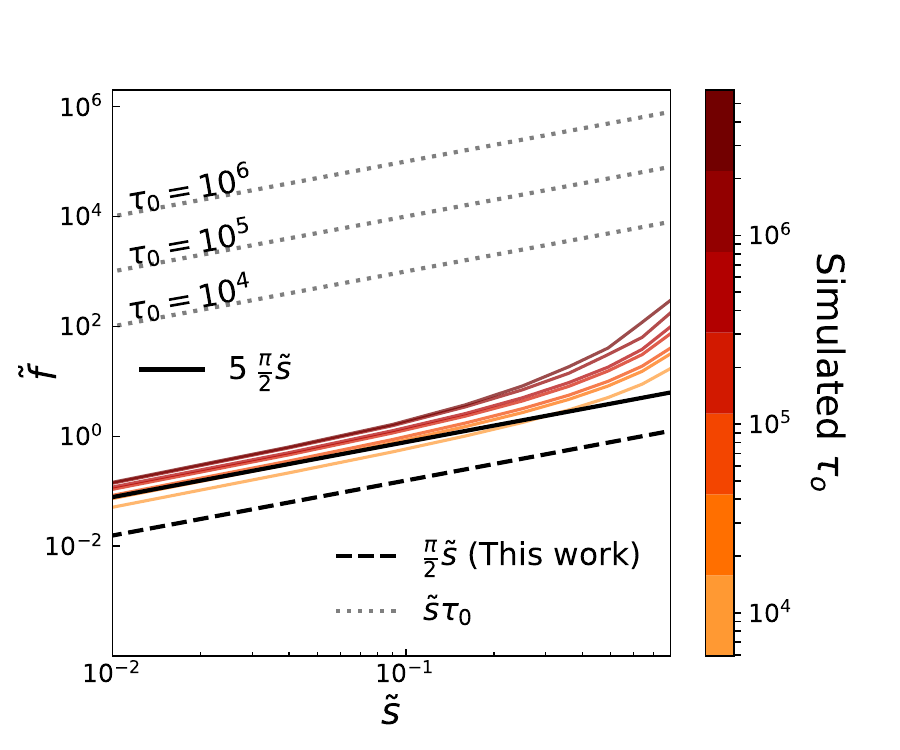}
    \vspace{-0.4cm}
    \caption{Hole-to slab ratio of escaping photons $\tilde f$ (described in \S~\ref{subsec:central_inconsistency}) as a function of the size of the hole and with varying optical depth.}
    \label{fig:centralvsS}
\end{figure}

Eq.~\ref{eq:ftilde2} establishes a linear relation between the hole-to-slab flux ratio  $\tilde f$ and the side area of the cavity $\tilde s$, independent of optical depth. In Fig.~\ref{fig:centralvsS}, the thin solid lines represent $\tilde f$ obtained from simulations for several values of $\tau_o$. Simulations replicate the $\sim \tilde s$ trend with a difference of a factor of $\sim 5$. We also include the previous theoretical prediction $\tilde f \sim \tilde s \tau_o$ for comparison. It is worth noting that both approaches differ by at least three orders of magnitude, meaning that the previous theoretical approach predicts that most of the flux will come from the hole rather than the slab. 

Fig.~\ref{fig:centralvsS} confirms that: (a) $\tilde f$ varies only slightly with optical depth, (b)  $\tilde f$ increases linearly with $\tilde s$ as expected by Eq.~\ref{eq:ftilde2}, and (c) contrary to previous expectations, the central peak intensity is comparable to the red and blue peaks. In other words, the presence of a large hole does not necessarily imply that all the flux will escape through it. 

The present work has some caveats. As mentioned earlier, both the anisotropic and the externally illuminated setups used for this study simplify the highly complicated gas structure in galaxies. We did not address additional factors that alter \Lya propagation, for instance, the presence of dust, known to absorb radiation and suppress the emergent \Lya line \citep{laursen_ly_2009}, or outflows, known to change the peaks velocity offset \citep{bonilha_monte_1979}. 
In future work, we will revisit these caveats and extend our model to different geometries, lines of sight, kinematics, dust content, and channel configurations. We will also discuss using these models to analyse observational data. 
Note also that the analytic model presented in \S~\ref{sec:new_approach} contains several simplifying assumptions. For instance, we assumed that the mean free path $\lambda_{\rm mfp}$ remains constant despite the frequency redistribution since we assume that photons mainly scatter in the line's core before escaping in a single excursion at $x\gtrsim x_{\rm esc}$. Furthermore, the factor $\alpha$ in Eq.~\ref{eq:Tslab} is a first-order approximation that depends on different optical depths\footnote{Note that with $\alpha=4/(3\tau(x_{\rm esc}))$ and $x_{\rm esc}=1.1*[a_v\tau_0/\sqrt{\pi}]^{1/3}$ \citep{harrington_scattering_1973, neufeld_transfer_1990}, $\tilde f$ shifts by a factor of $\sim 4.3$ for $\tau_0\simeq10^7$.}. These simplifications lead to the simulations fudge factor required to fully represent  Eq.~\ref{eq:ftilde2}. However, given the complexity of \Lya RT, our analytical model works surprisingly well and can explain the surprising escape behaviour observed.

\vspace{-0.5cm}
\section{Discussion \& Conclusion}
\label{sec:Discussion}

While gas distributions inside galaxies are well known to be highly anisotropic, there is still a lack of systematic studies on the resonant line transfer of \Lya through such geometries. This is highly surprising given that nature is generally anisotropic. In particular both theoretical \cite[e.g.,][]{mckee_theory_1977} and observational studies \citep[e.g.,][]{martin_mapping_2005} have well established that the gas distribution in the ISM is patchy with a wide variation in HI column densities. Motivated by this, we studied the simplest anisotropic setup: a HI-filled slab with a single, empty hole/channel carved in the neutral gas distribution. Regardless of the setup's simplicity, the results are counter-intuitive and far from simple. Given the considerable difference in line-centre optical depth between the slab and the hole (up to $\sim 10^7$), the naive initial expectation is that most \Lya photons escape through the empty channel, similar to light passing through the windows instead of the walls (see the discussion in \S~\ref{subsec:first_estimate}). Our numerical results show, in contrast, that only a fraction of photons escape through such a channel. 

To explain this dilemma, we developed an analytical model describing \Lya photons' general scattering and transmission behaviour. Our findings show that in an externally illuminated slab, photons projected towards the gas undergo a much greater number of scatterings before being reflected. In other words, reflected photons do not bounce out of the slab like ping-pong balls, but penetrate and scatter within the neutral gas before escaping from it. Specifically, we find that the number of scatterings per reflection follows a Lévy Distribution ($\propto N_{\rm scatter}^{-3/2}$ for large $N_{\rm scatter}$). These additional scatterings increase the chances of photons shifting their frequency enough to escape. As a consequence, these extra scatterings lead to a much greater transmission probability $\sim \tau_0^{-1/2}$ instead of the currently established and orders of magnitude smaller $\tau_0^{-1}$ \citep{neufeld_transfer_1990}. 

Our results imply that \Lya photons are surprisingly much less susceptible to the line-of-sight column densities than previously believed. In other words, \Lya photons don't probe only the path of `least resistance' (lowest column density) but rather a more general average of the distribution $p(N_{\rm HI})$. The new corrected transmission probability directly influences emergent \Lya spectra since photons trace information from more than one escape route instead of only the lowest density one.

In the setup of a slab with a hole, \Lya spectra trace photons, exiting both the hole (central peak) and the slab (red and blue peaks).
The altered picture of \Lya escape through anisotropic media can explain why triple-peaked spectra, like those in Fig.~\ref{fig:example_spec}, appear in our simulations instead of single-peaked spectra near the line centre. To quantify the \Lya radiation that escapes through the hole compared to the slab, we defined the hole-to-slab ratio in Eq.~\ref{eq:ftilde2}. We found that the central flux does not depend on the exact column density of the optically thick region but solely on the channel's size. Furthermore, as indicated by Fig.~\ref{fig:centralvsS}, channels with $\tilde s <0.06$ produce spectra with a central peak smaller than the blue and red peaks. Under these circumstances, the channel is potentially `hidden' even in a high column density system (see, for example, $\tilde s =0.01$ in the middle panel of Fig.~\ref{fig:example_spec}). Thus, \Lya could probe a porous ISM even while indicating high column densities with a wide peak separation. This implies, for instance, that observations of \Lya spectra pointing towards a high column density do not rule out the presence of low HI channels through which LyC can leak. 
This seems at odds with studies suggesting a clear correlation of the \Lya peak separation and LyC escape \citep[e.g.,][]{izotov_lyman_2021}. However, it might be that some of the (small, low-z) galaxies driving this correlation are rather isotropic systems and a population of anisotropic LyC leakers with larger \Lya peak separation exists. 

In summary, our study suggests that \Lya observables are shaped by average conditions rather than by only extremum statistics and can thus be used to study galaxy formation and evolution in a statistical manner.

\vspace{0.5cm}
\textbf{Acknowledgements.} We thank David Neufeld for the very useful discussions and encouragement and the anonymous referee for the very useful comments. SAM thanks the members of the Multiphase Gas at MPA, particularly Seok-Jun Chang and Hitesh Das, for the useful discussions and support. MG thanks the Max Planck Society for support through the Max Planck Research Group. This research was supported in part by grant NSF PHY-2309135 to the Kavli Institute for Theoretical Physics (KITP).

\vspace{0.5cm}

\textbf{Data Availability.} The data underlying this article will be shared on reasonable request to the corresponding author.
\bibliographystyle{mnras}
\bibliography{references.bib} 

\begin{thebibliography}{}
\makeatletter
\relax
\def\mn@urlcharsother{\let\do\@makeother \do\$\do\&\do\#\do\^\do\_\do\%\do\~}
\def\mn@doi{\begingroup\mn@urlcharsother \@ifnextchar [ {\mn@doi@} {\mn@doi@[]}}
\def\mn@doi@[#1]#2{\def\@tempa{#1}\ifx\@tempa\@empty \href {http://dx.doi.org/#2} {doi:#2}\else \href {http://dx.doi.org/#2} {#1}\fi \endgroup}
\def\mn@eprint#1#2{\mn@eprint@#1:#2::\@nil}
\def\mn@eprint@arXiv#1{\href {http://arxiv.org/abs/#1} {{\tt arXiv:#1}}}
\def\mn@eprint@dblp#1{\href {http://dblp.uni-trier.de/rec/bibtex/#1.xml} {dblp:#1}}
\def\mn@eprint@#1:#2:#3:#4\@nil{\def\@tempa {#1}\def\@tempb {#2}\def\@tempc {#3}\ifx \@tempc \@empty \let \@tempc \@tempb \let \@tempb \@tempa \fi \ifx \@tempb \@empty \def\@tempb {arXiv}\fi \@ifundefined {mn@eprint@\@tempb}{\@tempb:\@tempc}{\expandafter \expandafter \csname mn@eprint@\@tempb\endcsname \expandafter{\@tempc}}}

\bibitem[\protect\citeauthoryear{Adams}{Adams}{1972}]{adams_escape_1972}
Adams T.~F.,  1972, \mn@doi [The Astrophysical Journal] {10.1086/151503}, 174, 439

\bibitem[\protect\citeauthoryear{Ahn \& Lee}{Ahn \& Lee}{2002}]{ahn_ly_2002}
Ahn S.-H.,  Lee H.-W.,  2002, \mn@doi [Journal of Korean Astronomical Society] {10.5303/JKAS.2002.35.4.175}, 35, 175

\bibitem[\protect\citeauthoryear{Barnes, Garel  \& Kacprzak}{Barnes et~al.}{2014}]{barnes_ly-_2014}
Barnes L.~A.,  Garel T.,   Kacprzak G.~G.,  2014, \mn@doi [Publications of the Astronomical Society of the Pacific] {10.1086/679178}, 126, 969

\bibitem[\protect\citeauthoryear{Behrens \& Braun}{Behrens \& Braun}{2014}]{behrens_inclination_2014}
Behrens C.,  Braun H.,  2014, \mn@doi [Astronomy and Astrophysics] {10.1051/0004-6361/201424755}, 572, A74

\bibitem[\protect\citeauthoryear{Bik, Östlin, Menacho, Adamo, Hayes, Herenz  \& Melinder}{Bik et~al.}{2018}]{bik_super_2018}
Bik A.,  Östlin G.,  Menacho V.,  Adamo A.,  Hayes M.,  Herenz E.~C.,   Melinder J.,  2018, \mn@doi [Astronomy \& Astrophysics] {10.1051/0004-6361/201833916}, 619, A131

\bibitem[\protect\citeauthoryear{Bonilha, Ferch, Salpeter, Slater  \& Noerdlinger}{Bonilha et~al.}{1979}]{bonilha_monte_1979}
Bonilha J. R.~M.,  Ferch R.,  Salpeter E.~E.,  Slater G.,   Noerdlinger P.~D.,  1979, \mn@doi [The Astrophysical Journal] {10.1086/157426}, 233, 649

\bibitem[\protect\citeauthoryear{Chang, Arulanantham, Gronke, Herczeg  \& Bergin}{Chang et~al.}{2024}]{chang_first_2024}
Chang S.-J.,  Arulanantham N.,  Gronke M.,  Herczeg G.~J.,   Bergin E.~A.,  2024, \mn@doi [Monthly Notices of the Royal Astronomical Society] {10.1093/mnras/stae531}, 529, 2656

\bibitem[\protect\citeauthoryear{Cox \& Smith}{Cox \& Smith}{1974}]{cox_large-scale_1974}
Cox D.~P.,  Smith B.~W.,  1974, \mn@doi [The Astrophysical Journal] {10.1086/181476}, 189, L105

\bibitem[\protect\citeauthoryear{Dahle et~al.,}{Dahle et~al.}{2016}]{dahle_discovery_2016}
Dahle H.,  et~al., 2016, \mn@doi [Astronomy and Astrophysics] {10.1051/0004-6361/201628297}, 590, L4

\bibitem[\protect\citeauthoryear{Dijkstra}{Dijkstra}{2014}]{dijkstra_lyman_2014}
Dijkstra M.,  2014, \mn@doi [Publications of the Astronomical Society of Australia] {10.1017/pasa.2014.33}, 31, e040

\bibitem[\protect\citeauthoryear{Dijkstra, Haiman  \& Spaans}{Dijkstra et~al.}{2006}]{dijkstra_ly_2006}
Dijkstra M.,  Haiman Z.,   Spaans M.,  2006, \mn@doi [The Astrophysical Journal] {10.1086/506244}, 649, 37

\bibitem[\protect\citeauthoryear{Dijkstra, Gronke  \& Venkatesan}{Dijkstra et~al.}{2016}]{dijkstra_ly-lyc_2016}
Dijkstra M.,  Gronke M.,   Venkatesan A.,  2016, \mn@doi [The Astrophysical Journal] {10.3847/0004-637X/828/2/71}, 828, 71

\bibitem[\protect\citeauthoryear{Erb et~al.,}{Erb et~al.}{2014}]{erb_ly_2014}
Erb D.~K.,  et~al., 2014, \mn@doi [The Astrophysical Journal] {10.1088/0004-637X/795/1/33}, 795, 33

\bibitem[\protect\citeauthoryear{Gronke \& Dijkstra}{Gronke \& Dijkstra}{2014}]{gronke_directional_2014}
Gronke M.,  Dijkstra M.,  2014, \mn@doi [Monthly Notices of the Royal Astronomical Society] {10.1093/mnras/stu1513}, 444, 1095

\bibitem[\protect\citeauthoryear{Gronke \& Dijkstra}{Gronke \& Dijkstra}{2016}]{gronke_lyman-_2016}
Gronke M.,  Dijkstra M.,  2016, \mn@doi [The Astrophysical Journal] {10.3847/0004-637X/826/1/14}, 826, 14

\bibitem[\protect\citeauthoryear{Gronke, Dijkstra, McCourt  \& Oh}{Gronke et~al.}{2016}]{gronke_mirrors_2016}
Gronke M.,  Dijkstra M.,  McCourt M.,   Oh S.~P.,  2016, \mn@doi [The Astrophysical Journal] {10.3847/2041-8213/833/2/L26}, 833, L26

\bibitem[\protect\citeauthoryear{Gronke, Dijkstra, McCourt  \& Oh}{Gronke et~al.}{2017}]{gronke_resonant_2017}
Gronke M.,  Dijkstra M.,  McCourt M.,   Oh S.~P.,  2017, \mn@doi [Astronomy and Astrophysics] {10.1051/0004-6361/201731013}, 607, A71

\bibitem[\protect\citeauthoryear{Hansen \& Oh}{Hansen \& Oh}{2006}]{hansen_lyman_2006}
Hansen M.,  Oh S.~P.,  2006, \mn@doi [Monthly Notices of the Royal Astronomical Society] {10.1111/j.1365-2966.2005.09870.x}, 367, 979

\bibitem[\protect\citeauthoryear{Harrington}{Harrington}{1973}]{harrington_scattering_1973}
Harrington J.~P.,  1973, \mn@doi [Monthly Notices of the Royal Astronomical Society] {10.1093/mnras/162.1.43}, 162, 43

\bibitem[\protect\citeauthoryear{Hayes, Runnholm, Gronke  \& Scarlata}{Hayes et~al.}{2021}]{hayes_spectral_2021}
Hayes M.~J.,  Runnholm A.,  Gronke M.,   Scarlata C.,  2021, \mn@doi [The Astrophysical Journal] {10.3847/1538-4357/abd246}, 908, 36

\bibitem[\protect\citeauthoryear{Herenz, Hayes, Papaderos, Cannon, Bik, Melinder  \& Östlin}{Herenz et~al.}{2017}]{herenz_vltmuse_2017}
Herenz E.~C.,  Hayes M.,  Papaderos P.,  Cannon J.~M.,  Bik A.,  Melinder J.,   Östlin G.,  2017, \mn@doi [Astronomy and Astrophysics] {10.1051/0004-6361/201731809}, 606, L11

\bibitem[\protect\citeauthoryear{Izotov, Worseck, Schaerer, Guseva, Thuan, {Fricke}, Verhamme  \& Orlitová}{Izotov et~al.}{2018}]{izotov_low-redshift_2018}
Izotov Y.~I.,  Worseck G.,  Schaerer D.,  Guseva N.~G.,  Thuan T.~X.,  {Fricke} Verhamme A.,   Orlitová I.,  2018, \mn@doi [Monthly Notices of the Royal Astronomical Society] {10.1093/mnras/sty1378}, 478, 4851

\bibitem[\protect\citeauthoryear{Izotov, Worseck, Schaerer, Guseva, Chisholm, Thuan, Fricke  \& Verhamme}{Izotov et~al.}{2021}]{izotov_lyman_2021}
Izotov Y.~I.,  Worseck G.,  Schaerer D.,  Guseva N.~G.,  Chisholm J.,  Thuan T.~X.,  Fricke K.~J.,   Verhamme A.,  2021, \mn@doi [Monthly Notices of the Royal Astronomical Society] {10.1093/mnras/stab612}, 503, 1734

\bibitem[\protect\citeauthoryear{Jaskot, Dowd, Oey, Scarlata  \& McKinney}{Jaskot et~al.}{2019}]{jaskot_new_2019}
Jaskot A.~E.,  Dowd T.,  Oey M.~S.,  Scarlata C.,   McKinney J.,  2019, \mn@doi [The Astrophysical Journal] {10.3847/1538-4357/ab3d3b}, 885, 96

\bibitem[\protect\citeauthoryear{Kakiichi \& Gronke}{Kakiichi \& Gronke}{2021}]{kakiichi_lyman_2021}
Kakiichi K.,  Gronke M.,  2021, \mn@doi [The Astrophysical Journal] {10.3847/1538-4357/abc2d9}, 908, 30

\bibitem[\protect\citeauthoryear{Kim et~al.,}{Kim et~al.}{2023}]{kim_small_2023}
Kim K.~J.,  et~al., 2023, Small {Region}, {Big} {Impact}: {Highly} {Anisotropic} {Lyman}-continuum {Escape} from a {Compact} {Starburst} {Region} with {Extreme} {Physical} {Properties}, \url {http://arxiv.org/abs/2305.13405}

\bibitem[\protect\citeauthoryear{Laursen, Sommer-Larsen  \& Andersen}{Laursen et~al.}{2009}]{laursen_ly_2009}
Laursen P.,  Sommer-Larsen J.,   Andersen A.~C.,  2009, \mn@doi [The Astrophysical Journal] {10.1088/0004-637X/704/2/1640}, 704, 1640

\bibitem[\protect\citeauthoryear{Li \& Gronke}{Li \& Gronke}{2022}]{li_deciphering_2022}
Li Z.,  Gronke M.,  2022, \mn@doi [Monthly Notices of the Royal Astronomical Society] {10.1093/mnras/stac1207}, 513, 5034

\bibitem[\protect\citeauthoryear{Martin}{Martin}{2005}]{martin_mapping_2005}
Martin C.~L.,  2005, \mn@doi [The Astrophysical Journal] {10.1086/427277}, 621, 227

\bibitem[\protect\citeauthoryear{McKee \& Ostriker}{McKee \& Ostriker}{1977}]{mckee_theory_1977}
McKee C.~F.,  Ostriker J.~P.,  1977, \mn@doi [The Astrophysical Journal] {10.1086/155667}, 218, 148

\bibitem[\protect\citeauthoryear{Neufeld}{Neufeld}{1990}]{neufeld_transfer_1990}
Neufeld D.~A.,  1990, \mn@doi [The Astrophysical Journal] {10.1086/168375}, 350, 216

\bibitem[\protect\citeauthoryear{Neufeld}{Neufeld}{1991}]{neufeld_escape_1991}
Neufeld D.~A.,  1991, \mn@doi [The Astrophysical Journal] {10.1086/185983}, 370, L85

\bibitem[\protect\citeauthoryear{Orlitová, Verhamme, Henry, Scarlata, Jaskot, Oey  \& Schaerer}{Orlitová et~al.}{2018}]{orlitova_puzzling_2018}
Orlitová I.,  Verhamme A.,  Henry A.,  Scarlata C.,  Jaskot A.,  Oey M.~S.,   Schaerer D.,  2018, \mn@doi [Astronomy and Astrophysics] {10.1051/0004-6361/201732478}, 616, A60

\bibitem[\protect\citeauthoryear{Ouchi, Ono  \& Shibuya}{Ouchi et~al.}{2020}]{ouchi_observations_2020}
Ouchi M.,  Ono Y.,   Shibuya T.,  2020, Observations of the {Lyman}-\${\textbackslash}alpha\$ {Universe}, \mn@doi{10.1146/annurev-astro-032620-021859}, \url {https://arxiv.org/abs/2012.07960v1}

\bibitem[\protect\citeauthoryear{Rivera-Thorsen et~al.,}{Rivera-Thorsen et~al.}{2017}]{rivera-thorsen_sunburst_2017}
Rivera-Thorsen T.~E.,  et~al., 2017, \mn@doi [Astronomy and Astrophysics] {10.1051/0004-6361/201732173}, 608, L4

\bibitem[\protect\citeauthoryear{Rivera-Thorsen et~al.,}{Rivera-Thorsen et~al.}{2019}]{rivera-thorsen_gravitational_2019}
Rivera-Thorsen T.~E.,  et~al., 2019, \mn@doi [Science] {10.1126/science.aaw0978}, 366, 738

\bibitem[\protect\citeauthoryear{Smith et~al.,}{Smith et~al.}{2022}]{smith_physics_2022}
Smith A.,  et~al., 2022, \mn@doi [Monthly Notices of the Royal Astronomical Society] {10.1093/mnras/stac2641}, 517, 1

\bibitem[\protect\citeauthoryear{Steidel, Erb, Shapley, Pettini, Reddy, Bogosavljević, Rudie  \& Rakic}{Steidel et~al.}{2010}]{steidel_structure_2010}
Steidel C.~C.,  Erb D.~K.,  Shapley A.~E.,  Pettini M.,  Reddy N.,  Bogosavljević M.,  Rudie G.~C.,   Rakic O.,  2010, \mn@doi [The Astrophysical Journal] {10.1088/0004-637X/717/1/289}, 717, 289

\bibitem[\protect\citeauthoryear{Vanzella et~al.,}{Vanzella et~al.}{2020}]{vanzella_ionising_2020}
Vanzella E.,  et~al., 2020, \mn@doi [Monthly Notices of the Royal Astronomical Society] {10.1093/mnras/stz2286}, 491, 1093

\bibitem[\protect\citeauthoryear{Verhamme, Schaerer  \& Maselli}{Verhamme et~al.}{2006}]{verhamme_3d_2006}
Verhamme A.,  Schaerer D.,   Maselli A.,  2006, \mn@doi [Astronomy and Astrophysics] {10.1051/0004-6361:20065554}, 460, 397

\bibitem[\protect\citeauthoryear{Verhamme, Orlitová, Schaerer  \& Hayes}{Verhamme et~al.}{2015}]{verhamme_using_2015}
Verhamme A.,  Orlitová I.,  Schaerer D.,   Hayes M.,  2015, \mn@doi [Astronomy and Astrophysics] {10.1051/0004-6361/201423978}, 578, A7

\bibitem[\protect\citeauthoryear{Wisotzki et~al.,}{Wisotzki et~al.}{2016}]{wisotzki_extended_2016}
Wisotzki L.,  et~al., 2016, \mn@doi [Astronomy and Astrophysics] {10.1051/0004-6361/201527384}, 587, A98

\bibitem[\protect\citeauthoryear{Yang, Malhotra, Gronke, Rhoads, Dijkstra, Jaskot, Zheng  \& Wang}{Yang et~al.}{2016}]{yang_green_2016}
Yang H.,  Malhotra S.,  Gronke M.,  Rhoads J.~E.,  Dijkstra M.,  Jaskot A.,  Zheng Z.,   Wang J.,  2016, \mn@doi [The Astrophysical Journal] {10.3847/0004-637X/820/2/130}, 820, 130

\bibitem[\protect\citeauthoryear{Zackrisson, Inoue  \& Jensen}{Zackrisson et~al.}{2013}]{zackrisson_spectral_2013}
Zackrisson E.,  Inoue A.~K.,   Jensen H.,  2013, \mn@doi [The Astrophysical Journal] {10.1088/0004-637X/777/1/39}, 777, 39

\bibitem[\protect\citeauthoryear{Zheng \& Wallace}{Zheng \& Wallace}{2014}]{zheng_anisotropic_2014}
Zheng Z.,  Wallace J.,  2014, \mn@doi [The Astrophysical Journal] {10.1088/0004-637X/794/2/116}, 794, 116

\makeatother
\end{thebibliography}

\appendix
\section{Fraction of photons escaping without scattering}
\label{app:nointeraction}

\begin{figure}
    \centering
    \includegraphics[width=0.992\columnwidth, keepaspectratio]{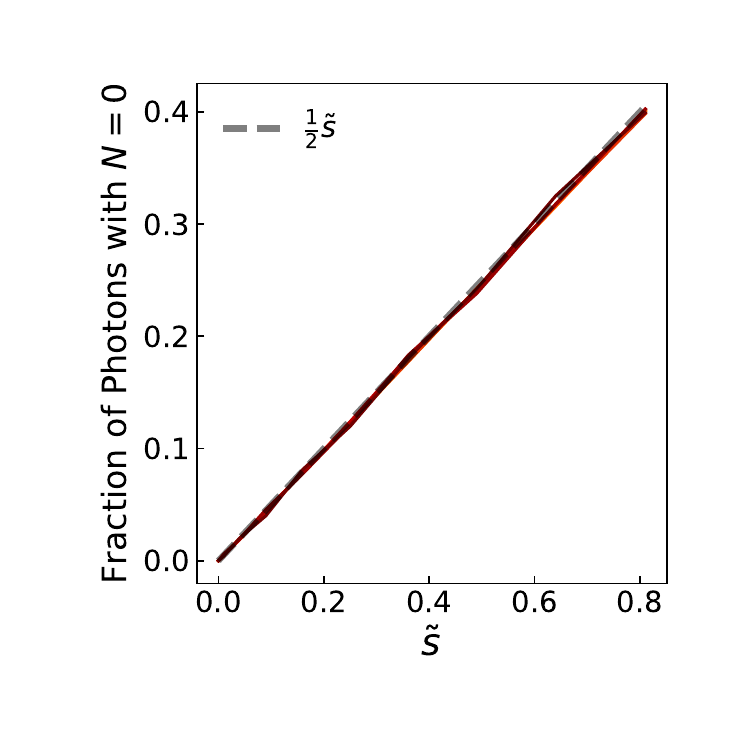}
    \vspace{-1cm}
    \caption{Fraction of photons that exit through the hole with no interaction with the gas ($N=0$). The colour coding corresponds to different optical depths as in Fig.~\ref{fig:centralvsS}.}
    \label{fig:nointeraction}
\end{figure}

Figure~\ref{fig:nointeraction} shows the fraction of photons undergoing no scatterings before escaping through the hole as a function of hole size. As expected, the fraction follows closely $\frac{\tilde s}{2}$ independent of the slab's column density. Note that the factor $1/2$ arises because of our choice of geometry and definition of $\tilde s$.

\bsp	
\label{lastpage}
\end{document}